\documentclass[%
superscriptaddress,
twocolumn,
nofootinbib,
amsmath,amssymb,
aps,
prd,
]{revtex4-2}

\usepackage[dvipsnames]{xcolor}
\usepackage[unicode]{hyperref}
\hypersetup{
  colorlinks=true,
  citecolor=MidnightBlue,
  linkcolor=MidnightBlue,
  urlcolor=MidnightBlue,
  linktocpage=true
}
\usepackage{tabularx}
\usepackage{graphicx}
\usepackage{amssymb,amsmath,bm,tensor,braket}
\usepackage[varg]{txfonts}
\usepackage{enumerate}
\usepackage{mathtools}
\usepackage{tensor}
\usepackage[capitalize]{cleveref}
\usepackage[utf8]{inputenc}
\usepackage[normalem]{ulem}
\usepackage{dcolumn}
\usepackage{soul}
\usepackage{subfigure}
\usepackage{booktabs}
\usepackage{comment}
\usepackage{siunitx}
\usepackage{mathrsfs}
\usepackage[T1]{fontenc}
\usepackage{float}

\newcommand{\dd}{\mathrm{d}}
\newcommand{\ii}{\mathrm{i}}

\newcounter{appslabel}
\renewcommand{\theappslabel}{\Alph{appslabel}}

\makeatletter
\newcommand{\appsection}[2]{%
  \section{#1}%
  \stepcounter{appslabel}%
  \edef\@currentlabel{\theappslabel}%
  \label{#2}%
}
\makeatother

\begin{document}

\title{Finite-Window Centered Organization of Neighboring Poles}

\author{Yuye Wu}
\affiliation{Department of Astronomy, College of Physical Science and Technology, Xiamen University, Xiamen 361005, China}

\author{Hong-Bo Jin}
\thanks{Corresponding author Email: \href{mailto:hbjin@bao.ac.cn}{hbjin@bao.ac.cn}}
\affiliation{National Astronomical Observatories, Chinese Academy of Sciences, Beijing 100101, China}
\affiliation{School of Astronomy and Space Science, University of Chinese Academy of Sciences, Beijing 100049, China}
\affiliation{The International Center for Theoretical Physics Asia-Pacific (ICTP-AP), University of Chinese Academy of Sciences, Beijing 100190, China}

\begin{abstract}
Near-degenerate resonance poles arise widely in open-wave systems. For gravitational-wave ringdowns, inference is performed on finite time windows where neighboring quasinormal modes can be spectrally close; the waveform is then dominated by a common carrier with a slowly varying interference envelope, while representing the signal as a sum of two independently resolved damped exponentials $e^{-\ii\omega_\pm t}$ becomes numerically ill-conditioned when the dimensionless splitting $\eta=|\sigma|T_{\mathrm{eff}}$ is small. We give a finite-window organizing principle for such neighboring-pole sectors: the local two-pole singular block of the Green-function integrand is rewritten exactly about a shared carrier $\omega_c$ and half-splitting $\sigma$, and for $|\sigma t|\ll 1$ the time-domain projection is systematically a carrier plus a first-jet piece $\propto t\,e^{-\ii\omega_c t}$, without requiring a literal double pole or exceptional-point merger in parameter space. The centered first-jet basis has $O(1)$ Gram conditioning, whereas the resolved-mode basis satisfies $\mathrm{cond}(G_{\mathrm{res}})\sim 12\,\eta^{-2}$ as $\eta\to 0$ (transparent real-splitting slice). We supply finite-window diagnostics in which $\kappa$ marks when the jet correction must be retained and $\eta^2$ sets the residual error scale once it is retained. Minimal two-pole numerics verify the scaling. For Kerr black holes we fix one adjacent-overtone mode pair (catalog label \texttt{pair45}; shared $(l,m)$ and consecutive overtones in our indexed tabulation), scan spin $a\in[0.8770,0.8810]$, and adopt the spectral window proxy $T_{\mathrm{spec}}=\beta/|\Im\omega_c|$ with $\beta=2.0$ to illustrate the same conditioning contrast in a near-degenerate sector.
\end{abstract}

\maketitle

%%%%%%%%%%%%%%%%%%%%%%%%%%%%%%%%
\noindent \textbf{\em Introduction.}
%%%%%%%%%%%%%%%%%%%%%%%%%%%%%%%%
Open-wave systems are governed by complex resonance poles. These poles move as parameters change. They can approach each other and become nearly degenerate. This near-degeneracy underlies many effects across optics, mesoscopic physics, and gravitational-wave ringdown. It also motivates the broader non-Hermitian language of exceptional points and resonance topology \cite{OzdemirEtAl2019, ElGanainyEtAl2018, RotterBird2015,Heiss2012,MiriAlu2019,BergholtzBudichKunst2021,AshidaGongUeda2020,PartoEtAl2021,DingFangMa2022,JaramilloPanossoMacedoAlSheikh2021,BhagwatEtAl2018StartTime}.

Kerr black holes provide a clean gravitational example. Their quasinormal modes (QNMs) can come in near-degenerate neighboring pairs. In enlarged parameter spaces, these pairs can show avoided-crossing-like behavior and exceptional-point structure \cite{YangEtAl2013,YangEtAl2013b,CavalcanteEtAl2024PRL,CavalcanteEtAl2024PRD}. Recent work also highlights the pair-dependent structure of nearby-overtone gaps, which sharpens the motivation to treat neighboring-pole sectors locally \cite{Wu:2026gyv}. Ringdown studies further highlight resonant excitation and destructive interference \cite{Motohashi2025,LoSabaniCardoso2025,YangBertiFranchini2025,OshitaBertiCardoso2025,MacedoEtAl2025}.

This paper focuses on a simple but practical point. Observations are made on a finite time window. On such a window, two close modes share an effective carrier frequency. The waveform then looks like a carrier multiplied by a slowly varying interference envelope. In this regime, fitting two independently resolved damped sinusoids can be numerically unstable. A more stable organization is needed.

Here we identify a finite-window response principle for near-degenerate neighboring poles. We work in a restricted local window where a neighboring pair dominates the response. In that setting, a centered description is more natural than a fully resolved two-pole description.

Concretely, we organize the local response as a centered two-pole block about a shared carrier. This is not just an algebraic rewrite. When the splitting is small, the resolved simple-pole basis becomes ill-conditioned on the observation window. In contrast, the centered first-jet basis remains regular. In the time domain, this corresponds to a carrier plus its first-jet (an associated-vector/Jordan-chain-like term), i.e., a carrier-plus-first-jet waveform.

This centered organization leads to a simple two-scale hierarchy on a finite window. The parameter $\kappa$ controls when the leading correction must be included. Once it is included, the remaining truncation error is controlled by $\eta^2$. Toy two-pole numerics verify this scaling, and Kerr quasinormal modes provide a representative gravitational realization.

For gravitational applications our aims are deliberately local: we characterize when the resolved-mode basis becomes ill-conditioned on a chosen window, construct the regular centered first-jet alternative, and express correction onset and truncation error in $(\kappa,\eta^2)$ in terms of $(\mathcal R_\pm,\sigma,T_{\mathrm{eff}})$. We do not replace global ringdown waveform models or prescribe detector-specific likelihoods; the results are a finite-window organizing principle to embed in reduced models or inference when a single neighboring pair dominates. Relative to exceptional-point studies in enlarged Kerr parameter spaces \cite{CavalcanteEtAl2024PRL,CavalcanteEtAl2024PRD}, which track degeneracy as external parameters vary, the emphasis here is the effective description on a fixed observation window when two poles are spectrally close---a complementary finite-time viewpoint on neighboring resonances.

\begin{figure*}[t]
\centering
\includegraphics[width=0.92\textwidth]{\detokenize{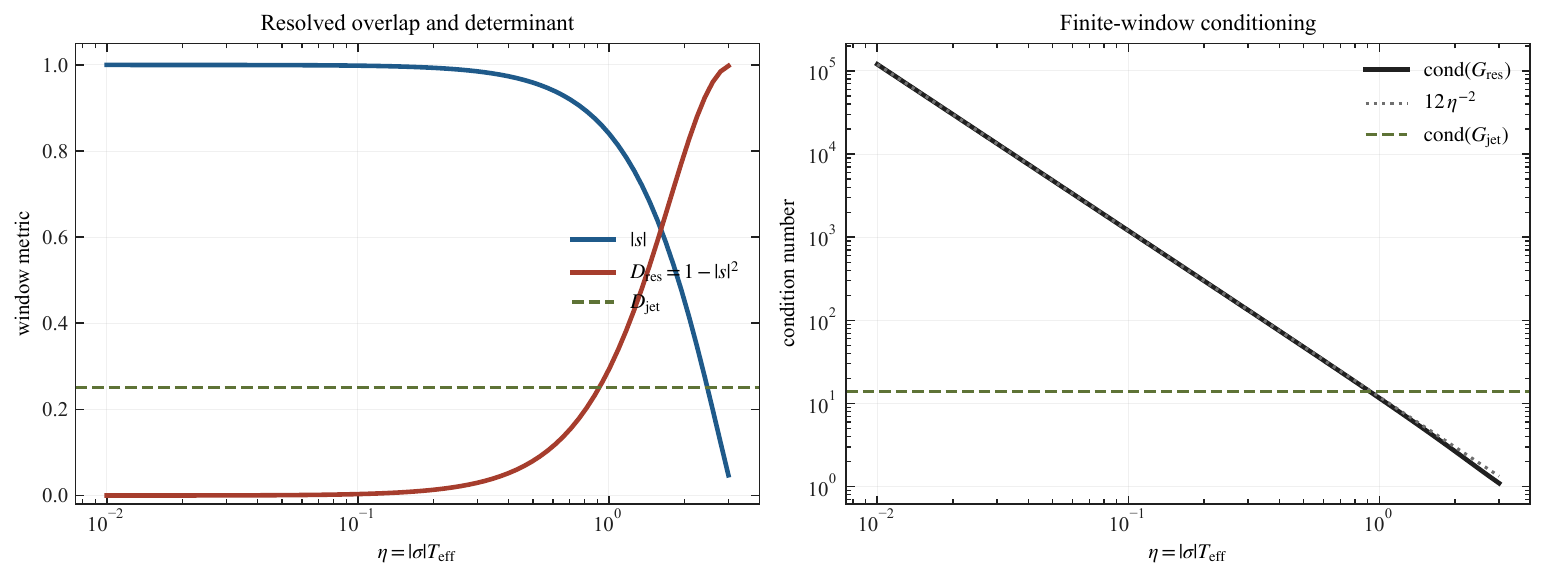}}
\caption{
Finite-window necessity of the centered organization.
Left: as $\eta=|\sigma|T_{\mathrm{eff}}\to 0$, the resolved basis $\{\phi_+,\phi_-\}$ becomes nearly linearly dependent on the observation window: the normalized overlap $s$ approaches unit modulus and the determinant proxy $D_{\mathrm{res}}=1-|s|^2$ collapses. By contrast, the centered first-jet basis retains an $O(1)$ determinant proxy.
Right: the resolved Gram matrix becomes ill-conditioned as $\mathrm{cond}(G_{\mathrm{res}})\sim 12\,\eta^{-2}$, whereas the centered first-jet Gram matrix remains $O(1)$, with $\mathrm{cond}(G_{\mathrm{jet}})=7+4\sqrt{3}\approx 13.93$. The centered organization is therefore not merely convenient: it is the regular low-order basis selected by the finite window.
}
\label{fig:necessity_conditioning}
\end{figure*}

%%%%%%%%%%%%%%%%%%%%%%%%%%%%%%%%
\noindent \textbf{\em Local effective two-pole response skeleton.}
%%%%%%%%%%%%%%%%%%%%%%%%%%%%%%%%
We start from the minimal object that controls the local response: a two-pole block in the Green-function integrand. This block captures a neighboring pair and their residues. The goal is to rewrite it in a form that makes the shared carrier and the small splitting explicit.
For ringdown, the relevant local object is the frequency-domain response integrand $\mathcal G(\omega;\lambda)$, as in the standard Green-function and quasinormal-mode formulation \cite{Chandrasekhar1985,Leaver1985,BertiCardosoStarinets2009}. Details of the derivation from the wave-equation Green function are given in Appendix~\ref{app:green}. In the standard quasinormal-mode picture, the singular part of $\mathcal G$ is organized as a sum of resolved simple poles, in parallel with the broader resonance-based description commonly used in open and non-Hermitian systems \cite{BertiCardosoStarinets2009,AshidaGongUeda2020,PartoEtAl2021}. The present problem concerns a restricted local window in which two neighboring poles dominate this singular structure. We therefore isolate the local two-pole block
\begin{equation}
    \mathcal G_{\mathrm {sing}}^{(2)}(\omega;\lambda)
    =
    \frac{\mathcal R_+(\lambda)}{\omega-\omega_+(\lambda)}
    +
    \frac{\mathcal R_-(\lambda)}{\omega-\omega_-(\lambda)}.
    \label{eq:two_pole_block}
\end{equation}
Here $\mathcal R_\pm$ are the local residues and $\omega_\pm$ are the neighboring pole positions. Following standard average-and-detuning coordinates in non-Hermitian two-mode systems \cite{ElGanainyEtAl2018,OzdemirEtAl2019}, we introduce the local center and half-splitting,
\begin{equation}
    \omega_c=\frac{\omega_++\omega_-}{2},
    \qquad
    \sigma=\frac{\omega_+-\omega_-}{2},
    \qquad
    x=\omega-\omega_c.
    \label{eq:center_split_green}
\end{equation}
Then Eq.~(\ref{eq:two_pole_block}) admits the exact rewrite
\begin{equation}
    \mathcal G_{\mathrm {sing}}^{(2)}(\omega;\lambda)
    =
    \frac{A(\lambda)\,x+B(\lambda)}{x^2-\sigma^2},
    \label{eq:centered_green_rewrite}
\end{equation}
with
\begin{equation}
    A(\lambda)=\mathcal R_+(\lambda)+\mathcal R_-(\lambda),
    \qquad
    B(\lambda)=\sigma\bigl[\mathcal R_+(\lambda)-\mathcal R_-(\lambda)\bigr].
    \label{eq:AB_def}
\end{equation}

Eq.~(\ref{eq:centered_green_rewrite}) is an exact local reorganization of the same neighboring-pole singular block. It does not assume a true double pole, nor an exact exceptional degeneracy in parameter space; rather, it isolates a local response block organized around a shared center and a small splitting, in a way that is conceptually adjacent to, but distinct from, the broader exceptional-point and non-Hermitian degeneracy framework \cite{BergholtzBudichKunst2021,DingFangMa2022}. Its role is local: on a restricted spectral window with neighboring-pole dominance, it identifies the minimal response block in terms of a shared center $\omega_c$, a local half-splitting $\sigma$, and the residue-weighted numerator combinations $A$ and $B$.

In this parametrization, $\omega_c$ specifies the common carrier of the local response block, while $\sigma$ measures the local spectral offset away from that carrier. The coefficients $A$ and $B$ encode how the two neighboring pole contributions combine within the same denominator structure. Eq.~(\ref{eq:centered_green_rewrite}) should therefore be viewed as the local analytic skeleton for the effective two-pole response.

The present framework is local in scope and intended for restricted windows with neighboring-pole dominance, excluding cases of anomalous zeroth-order suppression unless treated separately. The point is not that the simple-pole description disappears, but that, on such finite windows, the response is more naturally organized about a shared carrier than as two independently resolved contributions. In this sense, the centered form identifies a local regime of response reorganization rather than a literal pole merger.

%%%%%%%%%%%%%%%%%%%%%%%%%%%%%%%%
\noindent \textbf{\em Finite-window necessity of centered organization.}
%%%%%%%%%%%%%%%%%%%%%%%%%%%%%%%%
The centered organization is not only compact. It is also the numerically stable choice on a finite window. When the splitting is small, the two resolved exponentials become nearly linearly dependent on the window. The centered carrier-plus-first-jet basis stays well conditioned.
On a finite observation window, the centered organization is not merely an algebraic convenience. In the small-splitting regime, it is the regular low-order organization selected by the window itself. Consider the resolved basis
\begin{equation}
    \phi_\pm(t)=e^{-\,\ii(\omega_c\pm\sigma)t},
\end{equation}
and, as suggested by the two-pole Green-function contribution and its finite-window small-splitting expansion in Eqs.~(\ref{eq:two_mode_time_domain})--(\ref{eq:centered_block_effective_time_domain}) (see also Refs.~\cite{Leaver1985,BertiCardosoStarinets2009}), the centered first-jet basis
\begin{equation}
    \psi_0(t)=e^{-\,\ii\omega_c t},
    \qquad
    \psi_1(t)=t\,e^{-\,\ii\omega_c t},
\end{equation}
both restricted to the same interval $[0,T_{\mathrm{eff}}]$, and introduce the dimensionless parameter
\begin{equation}
    \eta = |\sigma|T_{\mathrm{eff}},
\end{equation}
motivated by finite-time windowing and resolvability considerations \cite{TalbotEtAl2021,ClarkeEtAl2024}.

As $\eta\to 0$, the two resolved exponentials become nearly linearly dependent on the observation window, whereas the centered carrier-plus-first-jet basis remains regular.

Along the transparent real-splitting slice, this distinction is explicit at the level of the normalized Gram matrices:
\begin{equation}
    \mathrm{cond}(G_{\mathrm{res}})=12\,\eta^{-2}+O(1),
    \qquad
    \mathrm{cond}(G_{\mathrm{jet}})=7+4\sqrt{3}\approx 13.93.
\end{equation}
Here the coefficient $12$ comes from the small-$\eta$ expansion of the normalized resolved-basis overlap, while $7+4\sqrt{3}$ is the exact condition number of the normalized centered first-jet Gram matrix; see Appendix~\ref{app:conditioning}.

Thus the resolved coordinates lose stability as the local splitting shrinks, while the centered coordinates remain $O(1)$. This finite-window contrast is shown directly in Fig.~\ref{fig:necessity_conditioning}: the resolved basis approaches linear dependence as $\eta\to 0$, whereas the centered first-jet basis retains an $O(1)$ determinant proxy and condition number. The point is therefore not merely that Eq.~(\ref{eq:centered_green_rewrite}) provides an exact rewrite, but that on a finite window it identifies the regular low-order basis, whereas the naively resolved basis becomes singular in practice.

\begin{figure*}[t]
\centering
\includegraphics[width=0.92\textwidth]{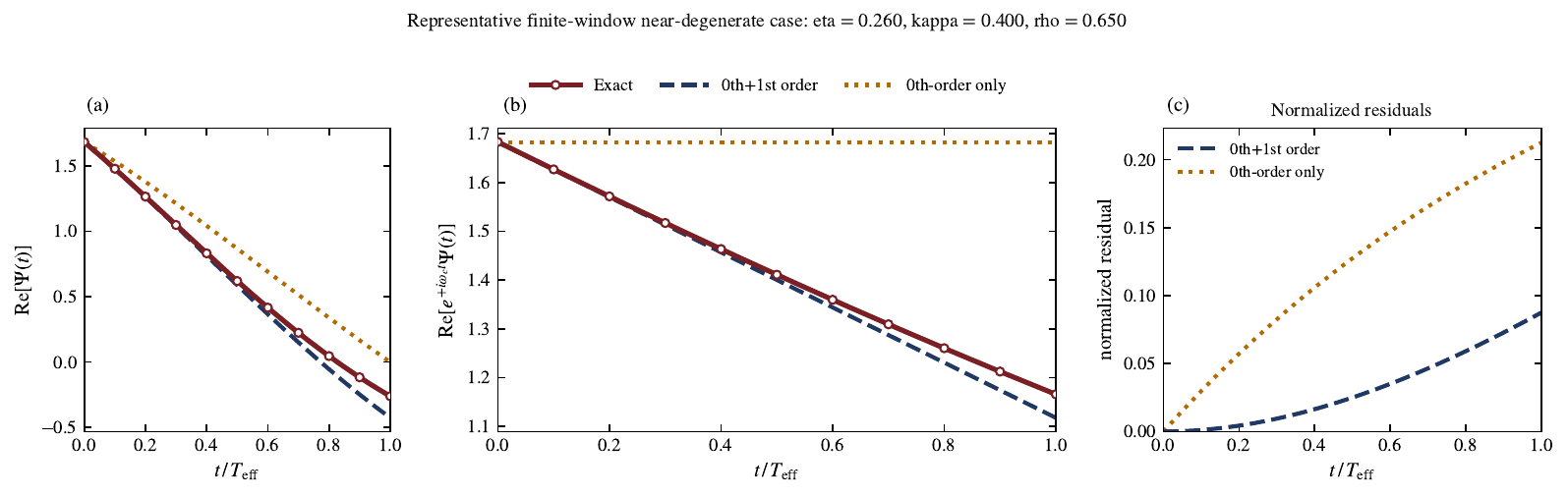}
\caption{
Waveform-level realization of the finite-window centered organization in a representative near-degenerate open-wave case. (a) Full waveform on the finite window. (b) After removing the common carrier $e^{-i\omega_c t}$, the reduced response is captured by the first-jet form, whereas the zeroth-order approximation misses the leading local drift. (c) Normalized residuals, showing that the first-order centered representation remains more accurate than the zeroth-order form. Parameters: $\eta=|\sigma|T_{\mathrm{eff}}=0.260$ and $\kappa=\eta\rho=0.400$, with $\rho=\left|(\mathcal R_+-\mathcal R_-)/(\mathcal R_++\mathcal R_-)\right|=0.650$.
}
\label{fig:openwave_time_landing}
\end{figure*}

This distinction is also confirmed numerically in coefficient-recovery tests; see Appendix~\ref{app:conditioning}. The centered organization is therefore selected both analytically, by basis regularity, and numerically, by inverse stability. Fig.~\ref{fig:openwave_time_landing} gives a representative waveform-level example.

%%%%%%%%%%%%%%%%%%%%%%%%%%%%%%%%
\noindent \textbf{\em Time-domain projection of local two-pole interference.}
%%%%%%%%%%%%%%%%%%%%%%%%%%%%%%%%
The same centered picture has a direct time-domain meaning. A near-degenerate pair produces a waveform with a common carrier and a slow interference envelope. On a short enough window, that envelope can be expanded. The result is a carrier plus its first-jet.
The finite-window necessity established above has a direct time-domain consequence. For the local centered two-pole block of Eq.~(\ref{eq:two_pole_block}), the pole contribution to the Green-function response familiar from quasinormal-mode theory \cite{Leaver1985,BertiCardosoStarinets2009} can be rewritten, after introducing the local center $\omega_c$ and half-splitting $\sigma$, as a common carrier multiplied by a splitting-dependent interference factor. On a finite window with $|\sigma|T_{\mathrm{eff}}\ll 1$, this factor admits a low-order expansion, yielding a carrier-plus-first-jet structure.

The two-pole contribution associated with Eq.~(\ref{eq:two_pole_block}) is
\begin{equation}
    \Psi^{(2)}(t;\lambda)
    \sim
    -\,\ii\Big[
    \mathcal R_+(\lambda)e^{-\,\ii\omega_+(\lambda)t}
    +
    \mathcal R_-(\lambda)e^{-\,\ii\omega_-(\lambda)t}
    \Big].
    \label{eq:two_mode_time_domain}
\end{equation}
Using Eq.~(\ref{eq:center_split_green}), this can be rewritten exactly as
\begin{equation}
    \Psi^{(2)}(t;\lambda)
    \sim
    -\,\ii\,e^{-\,\ii\omega_c t}
    \Big[
    (\mathcal R_++\mathcal R_-)\cos(\sigma t)
    -
    \ii(\mathcal R_+-\mathcal R_-)\sin(\sigma t)
    \Big].
    \label{eq:centered_time_domain_exact}
\end{equation}
This exact form already makes the centered organization explicit: the waveform is carried by the common complex frequency $\omega_c$, while the local splitting $\sigma$ enters only through the interference envelope around that shared carrier.

On a finite window satisfying $|\sigma t|\ll 1$, Eq.~(\ref{eq:centered_time_domain_exact}) reduces to
\begin{equation}
    \Psi^{(2)}(t;\lambda)
    \sim
    e^{-\,\ii\omega_c t}
    \bigl(C_0(\lambda)+C_1(\lambda)t+O(\sigma^2 t^2)\bigr),
    \label{eq:centered_block_effective_time_domain}
\end{equation}
where
\begin{equation}
    C_0(\lambda)=-\,\ii(\mathcal R_++\mathcal R_-),
    \qquad
    C_1(\lambda)=-(\mathcal R_+-\mathcal R_-)\sigma.
    \label{eq:C0C1_def}
\end{equation}
The carrier-plus-first-jet form in Eq.~(\ref{eq:centered_block_effective_time_domain}) is therefore not an auxiliary ansatz, but the finite-window time-domain projection of the same centered two-pole block identified on the frequency side.

Fig.~\ref{fig:openwave_time_landing} provides the waveform-level landing of Eq.~(\ref{eq:centered_block_effective_time_domain}) on a representative finite-window near-degenerate case. After removing the common carrier, the reduced waveform is accurately captured by the carrier-plus-first-jet form, while the corresponding residuals make the finite-window hierarchy explicit. In this sense, Fig.~\ref{fig:openwave_time_landing} is not an auxiliary illustration, but the time-domain realization of the same centered organization identified on the frequency side.

This expansion also exposes the local hierarchy needed below. The constant term gives the zeroth-order local description, the linear term gives the leading finite-window correction, and the remaining $O(\sigma^2 t^2)$ term sets the truncation scale of the corrected representation. The corresponding finite-window diagnostics therefore quantify not a separate mechanism, but the internal hierarchy of the same local two-pole response.

A residue-explicit Kerr implementation of these amplitude-side diagnostics is deferred here. At this stage, we establish only the time-domain realization of the finite-window centered organization itself; the later toy and Kerr analyses are used to test and illustrate its local control structure.

\begin{figure*}[t]
\centering
\includegraphics[width=0.44\textwidth]{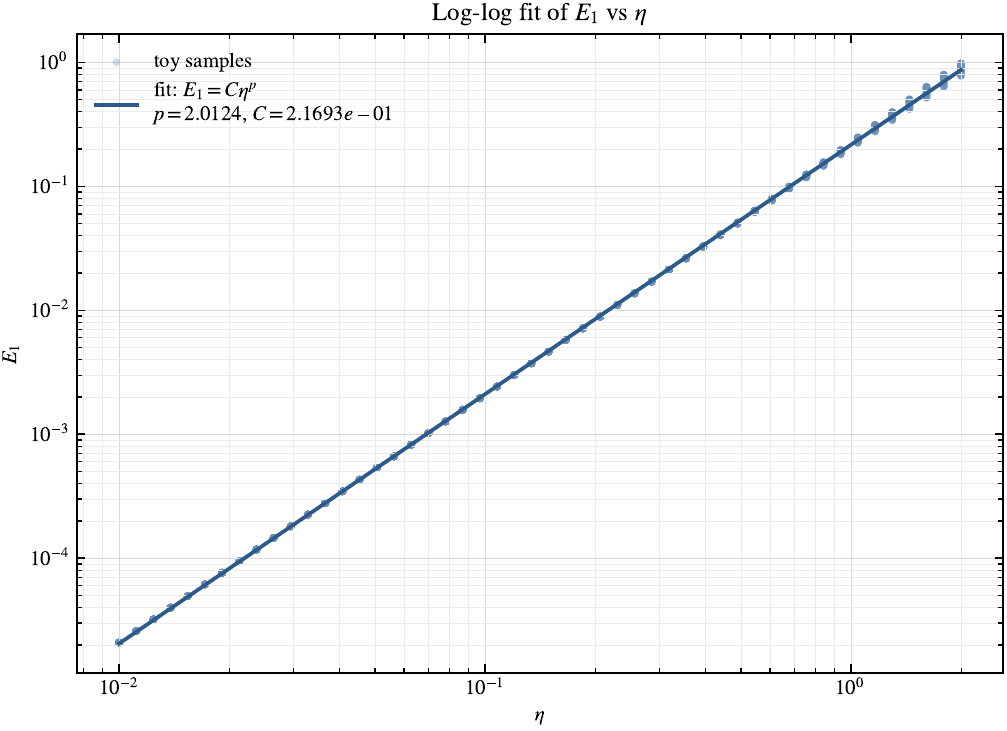}
\hfill
\includegraphics[width=0.44\textwidth]{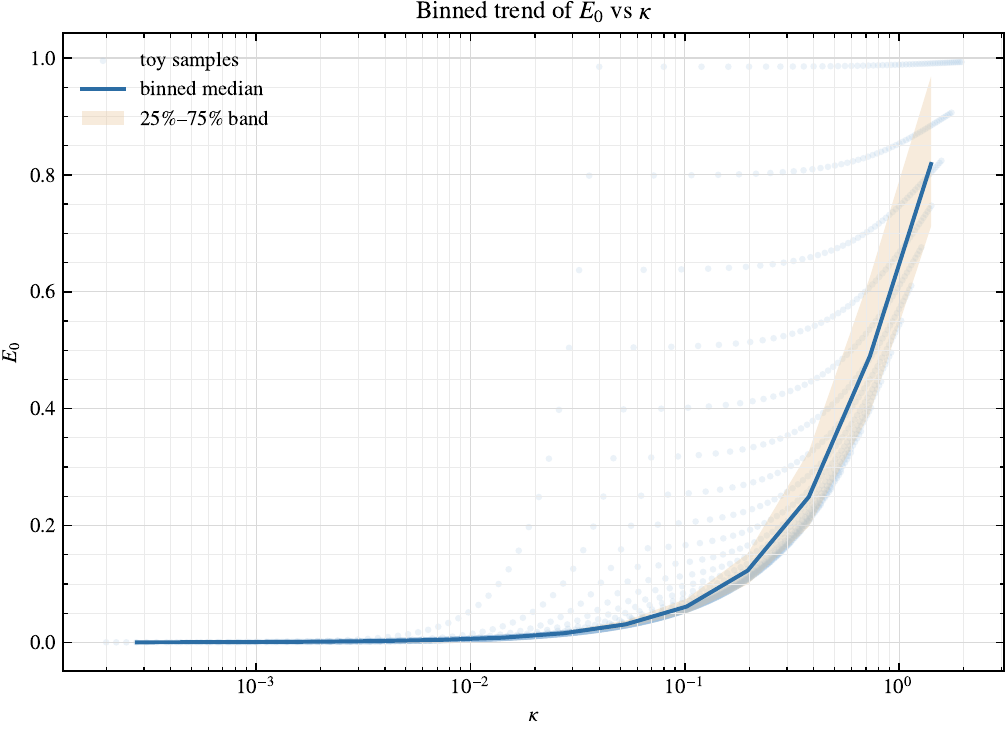}
\caption{
Toy two-pole numerics verify the finite-window two-scale hierarchy.
Left: the first-order residual error follows $E_1\propto \eta^{2.01}$, confirming that the corrected-representation error is governed by $\eta^2$.
Right: the zeroth-order error $E_0$ increases systematically with $\kappa$, showing that $\kappa$ controls the onset of the leading correction, while the remaining spread indicates subleading dependence beyond a single-$\kappa$ collapse.
}
\label{fig:toy_two_scale}
\end{figure*}

%%%%%%%%%%%%%%%%%%%%%%%%%%%%%%%%
\noindent \textbf{\em Local interference diagnostics.}
%%%%%%%%%%%%%%%%%%%%%%%%%%%%%%%%
The carrier-plus-first-jet expansion separates two practical questions. The first is when the linear correction is needed. The second is what sets the remaining truncation error once that correction is included. The parameters $\kappa$ and $\eta^2$ answer these two questions, respectively.
Eq.~(\ref{eq:centered_block_effective_time_domain}) immediately yields a finite-window hierarchy for the local two-pole response. Motivated by finite-time windowing and resolvability considerations in ringdown inference \cite{TalbotEtAl2021,ClarkeEtAl2024}, we quantify the relative importance of the linear correction on a window of size $T_{\mathrm{eff}}$ by
\begin{equation}
    \kappa
    =
    |\sigma|T_{\mathrm{eff}}
    \left|\frac{\mathcal R_+-\mathcal R_-}{\mathcal R_++\mathcal R_-}\right|
    =
    \eta\rho.
\end{equation}
For generic local windows in which the zeroth-order coefficient
$C_0\propto (\mathcal R_+ + \mathcal R_-)$ is not anomalously suppressed, $\kappa$ controls when the zeroth-order description ceases to be sufficient and the leading finite-window correction must be retained.

Once that linear correction is included, however, the controlling question changes. The remaining truncation error begins at $O(\sigma^2 t^2)$, so the residual accuracy of the corrected local representation is naturally governed at leading order by
\begin{equation}
    \eta^2=(|\sigma|T_{\mathrm{eff}})^2.
\end{equation}
The finite-window centered organization therefore has a genuine two-scale structure: $\kappa$ controls the onset of the leading correction, whereas $\eta^2$ controls the residual accuracy once that correction is retained.

The toy two-pole numerics verify this hierarchy directly. The first-order residual obeys an excellent power law $E_1\propto \eta^{2.01}$, confirming the expected $\eta^2$ scaling of the corrected-representation error. By contrast, the zeroth-order error $E_0$ is only partially organized by a single-$\kappa$ collapse, while two-scale fits involving both $\kappa$ and $\eta^2$ improve the description substantially. Fig.~\ref{fig:toy_two_scale} summarizes this structure.

The diagnostics therefore do not introduce a separate mechanism, but quantify two aspects of the same finite-window organization: $\kappa$ controls correction onset, whereas $\eta^2$ controls residual accuracy once that correction is retained.

\begin{figure}[t]
\centering
\includegraphics[width=0.98\columnwidth]{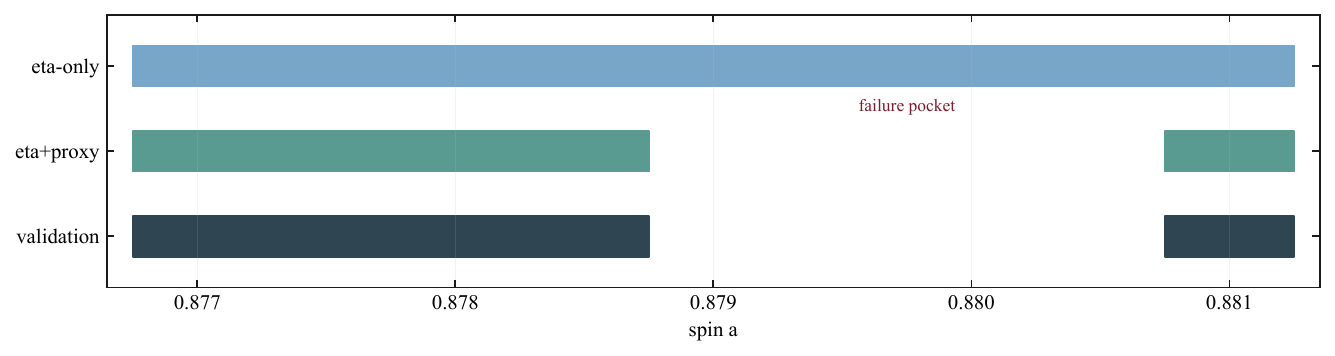}
\caption{
Local Kerr band comparison for the fixed adjacent-overtone pair \texttt{pair45} (shared $(l,m)$; consecutive overtones in our indexed Kerr table). Spin hull $a\in[0.8770,0.8810]$; $\eta_K$ uses $T_{\mathrm{spec}}=\beta/|\Im\omega_c|$ with $\beta=2.0$. The $\eta$-only band identifies a broad candidate region, the surrogate layer sharpens it, and the validation band marks the final locally supported subset (thresholds in Appendix~\ref{app:kerr}).
}
\label{fig:kerr_band_comparison}
\end{figure}

%%%%%%%%%%%%%%%%%%%%%%%%%%%%%%%%
\noindent \textbf{\em Representative Kerr realization.}
%%%%%%%%%%%%%%%%%%%%%%%%%%%%%%%%
Finally we test the same finite-window centered organization in a gravitational setting. We do not aim at a full amplitude-level ringdown model here. Instead we assess whether the same conditioning advantage of the centered basis appears for Kerr QNMs in a near-degenerate neighboring sector.
Kerr quasinormal modes furnish a representative gravitational realization of the finite-window centered organization identified above \cite{BertiCardosoStarinets2009,YangEtAl2013,CavalcanteEtAl2024PRL,CavalcanteEtAl2024PRD}. We hold fixed a single adjacent-overtone pair from our indexed Kerr catalogue, labeled \texttt{pair45}: both modes share the same angular quantum numbers $(l,m)$ and correspond to consecutive overtone indices in that tabulation (overtone ordering as in the underlying numerical tables \cite{BertiCardosoStarinets2009}). Denoting the pair frequencies by $\omega_\pm(a)$, we define
\begin{equation}
    \omega_c(a)=\frac{\omega_+(a)+\omega_-(a)}{2},
    \qquad
    \sigma(a)=\frac{\omega_+(a)-\omega_-(a)}{2}.
\end{equation}
To connect with finite-window diagnostics without fixing a detector-specific $T_{\mathrm{eff}}$, we use the spectral compressibility proxy from Appendix~\ref{app:kerr},
\begin{equation}
    T_{\mathrm{spec}}(a;\beta)=\frac{\beta}{|\Im \omega_c(a)|},
    \qquad
    \eta_K(a;\beta)=|\sigma(a)|\,T_{\mathrm{spec}}(a;\beta),
\end{equation}
with $\beta=2.0$ for the figures (scanning $\beta\in\{1.5,2.0,3.0\}$ rescales $\eta_K$ uniformly on the strip). The Kerr analysis is restricted: it does not supply a full amplitude-side implementation, but tests whether the same local reorganization appears in a nontrivial gravitational resonance problem. Fig.~\ref{fig:kerr_band_comparison} shows the spin hull $a\in[0.8770,0.8810]$ with the $\eta_K$-only candidate band, the surrogate sharpened layer, and the validated local support; all gate thresholds are recorded in Appendix~\ref{app:kerr}.

On this strip, the inverse problem reproduces the same finite-window distinction identified analytically above: throughout the small-$\eta$ regime, the resolved coefficient map exhibits substantially stronger perturbation amplification than the centered first-jet basis, with the contrast weakening as $\eta$ increases. This is most transparent at $a=0.8810$, for which $\eta=|\sigma|T_{\mathrm{spec}}=0.049145$ with $T_{\mathrm{spec}}=\beta/|\Im\omega_c|$ and $\beta=2.0$ (equivalently $\eta_K$ at the same $\beta$): the median and $q_{90}$ pair-spread ratios are $5.60$ and $7.35$, and the corresponding amplification ratio is $R_{\mathrm{amp}}=7.00$. These numbers show that, under the same perturbation level, the resolved inverse representation amplifies coefficient uncertainty by a factor of several relative to the centered one. We therefore use Kerr only in the restricted sense of a representative gravitational realization of the centered local response organization.

\noindent \textbf{\em Conclusion.}
We have identified a finite-window response principle for near-degenerate neighboring-pole sectors in open-wave systems, intended for restricted local windows with neighboring-pole dominance. On such a finite observation window, the resolved simple-pole basis becomes ill-conditioned as the local splitting shrinks, whereas the centered first-jet basis remains regular. The local response is therefore more naturally organized as a centered two-pole block about a shared carrier, with a corresponding carrier-plus-first-jet realization in the time domain. Within this organization, $\kappa$ controls correction onset, while $\eta^2$ controls the residual accuracy once that correction is retained. Toy two-pole numerics verify this two-scale hierarchy. For Kerr QNMs, fixing the adjacent-overtone pair \texttt{pair45} and scanning $a\in[0.8770,0.8810]$ with the spectral-window proxy $T_{\mathrm{spec}}=\beta/|\Im\omega_c|$ ($\beta=2.0$) illustrates the same finite-window organization in a near-degenerate sector, without claiming a complete ringdown amplitude model.

%%%%%%%%%%%%%%%%%%%%%%%%%%%%%%%%
\noindent \textbf{\em Acknowledgment.}
%%%%%%%%%%%%%%%%%%%%%%%%%%%%%%%%
This work is funded by the National Astronomical Observatories of the Chinese Academy of Sciences, Project No.E4TG6601. 
This work has been supported in part by the National Key Research and Development Program of China under Grant No.2021YFC2203000.

%%%%%%%%%%%%%%%%%%%%%%%%%%%%%%%%
\noindent \textbf{\em Data Availability.}
%%%%%%%%%%%%%%%%%%%%%%%%%%%%%%%%
Data can be obtained from the authors via reasonable request.

%%%%%%%%%%%%%%%%%%%%%%%%%%%%%%%%
\noindent \textbf{\em Competing Interests.}
%%%%%%%%%%%%%%%%%%%%%%%%%%%%%%%%
The authors have no competing financial or nonfinancial interests to disclose.

\clearpage
\appendix
\setcounter{appslabel}{0}
\makeatletter
\@removefromreset{equation}{section}
\setcounter{equation}{0}
\def\theequation@prefix{}%
\makeatother

%%%%%%%%%%%%%%%%%%%%%%%%%%%%%%%%
\noindent \textbf{\em Overview of appendices.}
%%%%%%%%%%%%%%%%%%%%%%%%%%%%%%%%
As in the main text, we work on a restricted local window with neighboring-pole dominance and keep the same notation for $\mathcal G(\omega;\lambda)$, $\omega_c$, $\sigma$, $\eta$, and $\kappa$. The main text states the local two-pole block and its centered rewrite, quotes the finite-window conditioning contrast and coefficient-recovery check, and summarizes the toy numerics and Kerr band criteria without repeating derivations. The following appendices collect the corresponding details in the same order as they support the narrative above: from the wave-equation Green function to the effective integrand $\mathcal G(\omega;\lambda)$ (the skeleton of the ``Local effective two-pole response'' discussion), then to the time-domain carrier-plus-first-jet expansion and $\kappa$--$\eta^2$ hierarchy (underlying ``Time-domain projection'' and ``Local interference diagnostics''), then to the Gram-matrix and recovery analysis behind ``Finite-window necessity,'' the controlled two-pole protocol behind Fig.~\ref{fig:toy_two_scale}, and finally the Kerr thresholds behind Fig.~\ref{fig:kerr_band_comparison}.

\appsection{Green-function route to the local two-pole block}{app:green}

We start, as usual in black-hole perturbation theory \cite{Chandrasekhar1985, Leaver19861, Leaver19862, BertiCardosoStarinets2009, Teukolsky1973}, from the master wave equation and its Green-function formulation \cite{Vishveshwara1970,Press1971}, and reduce the construction to the frequency-domain object that actually enters the local pole bookkeeping in the main text.
\begin{equation}
    \left[
    \partial_t^2-\partial_{r_*}^2+V_\lambda(r_*)
    \right]
    \Psi(t,r_*)=0,
\end{equation}
where $r_*$ is the tortoise coordinate and $\lambda$ denotes an external control parameter.

The retarded Green function for this problem satisfies
\begin{equation}
    \left[
    \partial_t^2-\partial_{r_*}^2+V_\lambda(r_*)
    \right]
    G_{\mathrm{ret}}(t;r_*,r_*';\lambda)
    =
    \delta(t)\delta(r_*-r_*')
\end{equation}
with
\begin{equation}
    G_{\mathrm{ret}}(t;r_*,r_*';\lambda)=0,
    \qquad t<0.
    \label{eq:S_retarded_condition}
\end{equation}
Passing to the frequency domain, we write
\begin{equation}
    G_{\mathrm{ret}}(t;r_*,r_*';\lambda)
    =
    \int_{\Gamma_\omega}
    \frac{\dd\omega}{2\pi}\,
    e^{-\,\ii\omega t}\,
    \widetilde G(\omega;r_*,r_*';\lambda),
    \qquad t>0,
    \label{eq:S_Gret_fourier}
\end{equation}
where the contour $\Gamma_\omega$ is chosen consistently with retarded evolution. The frequency-domain Green function obeys
\begin{equation}
    \left[
    -\partial_{r_*}^2+V_\lambda(r_*)-\omega^2
    \right]
    \widetilde G(\omega;r_*,r_*';\lambda)
    =
    \delta(r_*-r_*').
    \label{eq:S_freq_green_eq}
\end{equation}

Let $u_{\rm in}(\omega,r_*;\lambda)$ and $u_{\rm up}(\omega,r_*;\lambda)$ denote incoming and upgoing solutions satisfying the standard asymptotic boundary conditions in the Kerr/black-hole perturbation construction \cite{Leaver1985,Leaver19862}. In terms of these modes,
\begin{equation}
    \widetilde G(\omega;r_*,r_*';\lambda)
    =
    \frac{
    u_{\mathrm{in}}(\omega,r_<;\lambda)\,
    u_{\mathrm{up}}(\omega,r_>;\lambda)
    }{
    W(\omega;\lambda)
    },
    \label{eq:S_freq_green_wronskian}
\end{equation}
where $r_< =\min(r_*,r_*')$, $r_> = \max(r_*,r_*')$, and
\begin{equation}
    W(\omega;\lambda)
    =
    u_{\mathrm{in}}\,\partial_{r_*}u_{\mathrm{up}}
    -
    u_{\mathrm{up}}\,\partial_{r_*}u_{\mathrm{in}}
    \label{eq:S_wronskian_def}
\end{equation}
is the Wronskian. Quasinormal poles correspond to zeros of $W(\omega;\lambda)$, as in the standard Green-function and spectral-decomposition viewpoint \cite{Leaver1985,Leaver19861}.

For ringdown observed at $r_*^{\rm obs}$, with initial data encoded by a frequency-domain source functional $\mathcal I(\omega,r_*')$, the time-domain waveform admits the schematic representation
\begin{equation}
    \Psi(t,r_*^{\rm obs};\lambda)
    =
    \int_{\Gamma_\omega}\frac{\dd\omega}{2\pi}\,
    e^{-\,\ii\omega t}\,
    \mathcal G(\omega;\lambda),
    \label{eq:S_waveform_green_rep}
\end{equation}
with
\begin{equation}
    \mathcal G(\omega;\lambda)
    =
    \int \dd r_*'\,
    \widetilde G(\omega;r_*^{\rm obs},r_*';\lambda)\,
    \mathcal I(\omega,r_*').
    \label{eq:S_effective_green_integrand}
\end{equation}
The local pole organization emphasized in the main text is that of the integrand $\mathcal G(\omega;\lambda)$, not the full late-time Green-function anatomy, which can also carry non-pole contributions \cite{Price1972,ChingEtAl1995,Leaver19861}.

For the present local construction, we retain only the neighboring-pole sector of the response. In the standard quasinormal-mode picture, the singular part of $\mathcal G(\omega;\lambda)$ is organized as a sum of simple poles,
\begin{equation}
    \mathcal G(\omega;\lambda)\sim
    \frac{\mathcal R_n(\lambda)}{\omega-\omega_n(\lambda)}.
    \label{eq:S_simple_pole_local}
\end{equation}
On a spectral window where two neighboring poles dominate, we isolate the same two-pole block as in the main text,
\begin{equation}
    \mathcal G_{\mathrm{sing}}^{(2)}(\omega;\lambda)
    =
    \frac{\mathcal R_+(\lambda)}{\omega-\omega_+(\lambda)}
    +
    \frac{\mathcal R_-(\lambda)}{\omega-\omega_-(\lambda)}.
    \label{eq:S_two_pole_block}
\end{equation}

Introducing the average-and-detuning coordinates already fixed in the main text,
\begin{equation}
    \omega_c=\frac{\omega_++\omega_-}{2},
    \qquad
    \sigma=\frac{\omega_+-\omega_-}{2},
    \qquad
    x=\omega-\omega_c,
    \label{eq:S_center_split_green}
\end{equation}
one has
\begin{equation}
    \omega-\omega_+=x-\sigma,
    \qquad
    \omega-\omega_-=x+\sigma.
    \label{eq:S_shifted_denominators}
\end{equation}
Combining denominators yields
\begin{align}
    \mathcal G_{\mathrm{sing}}^{(2)}(\omega;\lambda)
    &=
    \frac{\mathcal R_+}{x-\sigma}
    +
    \frac{\mathcal R_-}{x+\sigma}
    \nonumber\\
    &=
    \frac{
    \mathcal R_+(x+\sigma)+\mathcal R_-(x-\sigma)
    }{
    (x-\sigma)(x+\sigma)
    }
    \nonumber\\
    &=
    \frac{
    (\mathcal R_++\mathcal R_-)x
    +
    \sigma(\mathcal R_+-\mathcal R_-)
    }{
    x^2-\sigma^2
    }.
    \label{eq:S_centered_rewrite_derivation}
\end{align}
which is the centered form quoted in the main text,
\begin{equation}
    \mathcal G_{\mathrm{sing}}^{(2)}(\omega;\lambda)
    =
    \frac{A(\lambda)\,x+B(\lambda)}{x^2-\sigma^2},
    \label{eq:S_centered_green_rewrite}
\end{equation}
with coefficients
\begin{equation}
    A(\lambda)=\mathcal R_+(\lambda)+\mathcal R_-(\lambda),
    \qquad
    B(\lambda)=\sigma\bigl[\mathcal R_+(\lambda)-\mathcal R_-(\lambda)\bigr].
    \label{eq:S_AB_def}
\end{equation}

\appsection{Finite-window time-domain expansion and amplitude hierarchy}{app:time}

This section spells out the time-domain transcription of the local two-pole block and the small-$|\sigma t|$ expansion that underlies the carrier-plus-first-jet form and the $\kappa$--$\eta^2$ diagnostics in the main text. In the standard quasinormal-mode response picture \cite{Press1971,Leaver19861,Andersson1995}, the pole contribution associated with \eqref{eq:S_two_pole_block} reads
\begin{equation}
    \Psi^{(2)}(t;\lambda)
    \sim
    -\,\ii\Big[
    \mathcal R_+(\lambda)e^{-\,\ii\omega_+(\lambda)t}
    +
    \mathcal R_-(\lambda)e^{-\,\ii\omega_-(\lambda)t}
    \Big].
    \label{eq:S_two_mode_time_domain}
\end{equation}
With $\omega_\pm=\omega_c\pm\sigma$, the common carrier $e^{-\,\ii\omega_c t}$ can be factored out explicitly:
\begin{equation}
    \Psi^{(2)}(t;\lambda)
    \sim
    -\,\ii\,e^{-\,\ii\omega_c t}
    \Big[
    \mathcal R_+ e^{-\,\ii\sigma t}
    +
    \mathcal R_- e^{\,\ii\sigma t}
    \Big].
    \label{eq:S_factor_common_carrier}
\end{equation}
Using Euler's formula,
\begin{equation}
    e^{\pm \ii\sigma t}=\cos(\sigma t)\pm \ii\sin(\sigma t),
    \label{eq:S_euler}
\end{equation}
one obtains the centered interference form
\begin{equation}
    \Psi^{(2)}(t;\lambda)
    \sim
    -\,\ii\, e^{-\,\ii\omega_c t}
    \Big[
    (\mathcal R_+ + \mathcal R_-)\cos(\sigma t)
    -
    \ii(\mathcal R_+-\mathcal R_-)\sin(\sigma t)
    \Big].
    \label{eq:S_centered_time_domain_exact}
\end{equation}

On a finite observation window with $|\sigma t|\ll 1$, we may use
\begin{equation}
    \cos(\sigma t)=1+O(\sigma^2 t^2),
    \qquad
    \sin(\sigma t)=\sigma t+O(\sigma^3 t^3),
    \label{eq:S_small_sigma_t}
\end{equation}
and expand to obtain
\begin{align}
    \Psi^{(2)}(t;\lambda)
    &\sim
    -\,\ii\,e^{-\,\ii\omega_c t}
    \Big[
    (\mathcal R_++\mathcal R_-)
    -
    \ii(\mathcal R_+-\mathcal R_-)\sigma t
    +
    O(\sigma^2 t^2)
    \Big].
    \label{eq:S_centered_time_domain_expansion}
\end{align}
Equivalently,
\begin{equation}
    \Psi^{(2)}(t;\lambda)
    \sim
    e^{-\,\ii\omega_c t}
    \bigl(C_0(\lambda)+C_1(\lambda)t+O(\sigma^2 t^2)\bigr),
    \label{eq:S_centered_block_effective_time_domain}
\end{equation}
where
\begin{equation}
    C_0(\lambda)=-\,\ii(\mathcal R_++\mathcal R_-),
    \qquad
    C_1(\lambda)=-(\mathcal R_+-\mathcal R_-)\sigma.
    \label{eq:S_C0C1_def}
\end{equation}

On a window of size $T_{\mathrm{eff}}$, the relative importance of the linear correction is captured by the same dimensionless combination as in the main text,
\begin{equation}
    \kappa
    =
    |\sigma|T_{\mathrm{eff}}
    \left|
    \frac{\mathcal R_+ - \mathcal R_-}{\mathcal R_+ + \mathcal R_-}
    \right|.
    \label{eq:S_kappa_def}
\end{equation}
Away from accidental suppression of $C_0\propto(\mathcal R_++\mathcal R_-)$, $\kappa$ marks when the zeroth-order carrier description ceases to be adequate and the leading finite-window correction must be kept. Once that correction is retained, the residual error starts at $O(\sigma^2 t^2)$, so the accuracy of the corrected representation is governed at leading order by
\begin{equation}
    \eta^2=(|\sigma|T_{\mathrm{eff}})^2.
    \label{eq:S_eta2_def}
\end{equation}

\appsection{Conditioning derivation and coefficient-recovery test}{app:conditioning}

The conditioning estimates cited in the main text follow from elementary Gram-matrix calculations on the finite window. Consider the resolved basis
\begin{equation}
    \phi_\pm(t)=e^{-\,\ii(\omega_c\pm\sigma)t},
\end{equation}
and the centered first-jet pair
\begin{equation}
    \psi_0(t)=e^{-\,\ii\omega_c t},
    \qquad
    \psi_1(t)=t\,e^{-\,\ii\omega_c t},
\end{equation}
both restricted to the same interval $[0,T_{\mathrm{eff}}]$ as in the main text. The dimensionless splitting parameter is again
\begin{equation}
    \eta=|\sigma|T_{\mathrm{eff}},
\end{equation}
motivated by the same finite-time windowing and resolvability considerations as in ringdown inference \cite{TalbotEtAl2021,ClarkeEtAl2024}. For the Gram-matrix bookkeeping we employ the normalized overlap
\begin{equation}
    \langle f,g\rangle_T
    =
    \frac{\int_0^{T_{\mathrm{eff}}} f^*(t)\,g(t)\,\dd t}
    {\|f\|\,\|g\|}.
\end{equation}

For the resolved exponentials, the normalized overlap is
\begin{equation}
    s=\langle \phi_+,\phi_-\rangle_T.
\end{equation}
Along the transparent real-splitting slice this reduces to
\begin{equation}
    |s|=\frac{|\sin\eta|}{\eta}
    =
    1-\frac{\eta^2}{6}+O(\eta^4),
\end{equation}
so that
\begin{equation}
    D_{\mathrm{res}}
    =
    1-|s|^2
    =
    \frac{\eta^2}{3}+O(\eta^4).
\end{equation}
In particular, the normalized resolved Gram matrix becomes ill-conditioned as
\begin{equation}
    \mathrm{cond}(G_{\mathrm{res}})
    =
    \frac{1+|s|}{1-|s|}
    =
    12\,\eta^{-2}+O(1),
    \qquad \eta\to 0.
\end{equation}

By contrast, the centered first-jet pair yields an $\eta$-independent normalized Gram matrix with
\begin{equation}
    |s_{\mathrm{jet}}|=\frac{\sqrt{3}}{2},
    \qquad
    D_{\mathrm{jet}}=\frac14,
    \qquad
    \mathrm{cond}(G_{\mathrm{jet}})=7+4\sqrt{3}\approx 13.93.
\end{equation}
Thus the resolved representation loses stable coordinates when the splitting is small, whereas the centered first-jet coordinates remain $O(1)$ on the same window---the analytic content behind Fig.~\ref{fig:necessity_conditioning} in the main text.

The same contrast shows up in a minimal coefficient-recovery test. With additive complex Gaussian noise on synthetic two-mode waveforms, reconstructions in the resolved basis amplify coefficient uncertainty far more strongly than reconstructions in the centered first-jet basis when $\eta$ is small. At noise level $\epsilon=10^{-4}$, the median coefficient instability is $8.26\times10^{-4}$ versus $9.94\times10^{-6}$ at $\eta=10^{-2}$, and $9.16\times10^{-5}$ versus $1.11\times10^{-5}$ at $\eta=10^{-1}$; the two organizations become comparable only when $\eta$ approaches unity.

\appsection{Toy-model protocol and fitting details}{app:toy}

The toy panel in the main text is produced from a deliberately minimal protocol: we work directly in the local two-pole class, without additional physics, so that the scaling of $E_0$ and $E_1$ isolates the finite-window hierarchy itself.

%%%%%%%%%%%%%%%%%%%%%%%%%%%%%%%%
\noindent \textbf{\em Toy setup.}
%%%%%%%%%%%%%%%%%%%%%%%%%%%%%%%%
To stress-test the $\kappa$--$\eta^2$ picture developed in Appendix~\ref{app:time}, we generate synthetic waveforms from the exact resolved two-pole form. The reference signal is
\begin{equation}
    \Psi_{\mathrm{exact}}(t)= -\,\ii\Big[R_+ e^{-\,\ii(\omega_c+\sigma)t}+R_- e^{-\,\ii(\omega_c-\sigma)t}\Big],
\end{equation}
with fixed
\begin{equation}
    \omega_c = 1.0 - 0.2\,\ii,
    \qquad
    T_{\mathrm{eff}}=1.0,
\end{equation}
and a uniform grid of 4000 samples on $t\in[0,T_{\mathrm{eff}}]$.

The zeroth- and first-order local approximants are the carrier-only and carrier-plus-first-jet truncations,
\begin{equation}
    \Psi^{(0)}(t)= -\,\ii\,e^{-\,\ii\omega_c t}(R_+ + R_-),
\end{equation}
\begin{equation}
    \Psi^{(1)}(t)= e^{-\,\ii\omega_c t}\Big[-\,\ii(R_+ + R_-) - (R_+ - R_-)\sigma t\Big].
\end{equation}

Samples are drawn on an $(\eta,\rho)$ grid with
\begin{equation}
    \eta = |\sigma|T_{\mathrm{eff}}\in [10^{-2},10^{0.3}],
    \qquad
    \rho\in[0,0.98],
\end{equation}
using 50 logarithmically spaced values of $\eta$ and 50 linearly spaced values of $\rho$, for a total of $N=2500$ samples. The residues are chosen as
\begin{equation}
    R_+ = \frac{1+\rho}{2},
    \qquad
    R_- = \frac{1-\rho}{2},
\end{equation}
so that
\begin{equation}
    \kappa = \eta \left|\frac{R_+-R_-}{R_++R_-}\right| = \eta\rho .
\end{equation}

%%%%%%%%%%%%%%%%%%%%%%%%%%%%%%%%
\noindent \textbf{\em Error measures.}
%%%%%%%%%%%%%%%%%%%%%%%%%%%%%%%%
Relative $L^2$ errors on $t\in[0,T_{\mathrm{eff}}]$ are
\begin{equation}
    E_0 =
    \left(
    \frac{\int_0^{T_{\mathrm{eff}}} |\Psi_{\mathrm{exact}}(t)-\Psi^{(0)}(t)|^2\,\dd t}
    {\int_0^{T_{\mathrm{eff}}} |\Psi_{\mathrm{exact}}(t)|^2\,\dd t}
    \right)^{1/2},
\end{equation}
\begin{equation}
    E_1 =
    \left(
    \frac{\int_0^{T_{\mathrm{eff}}} |\Psi_{\mathrm{exact}}(t)-\Psi^{(1)}(t)|^2\,\dd t}
    {\int_0^{T_{\mathrm{eff}}} |\Psi_{\mathrm{exact}}(t)|^2\,\dd t}
    \right)^{1/2}.
\end{equation}

%%%%%%%%%%%%%%%%%%%%%%%%%%%%%%%%
\noindent \textbf{\em Fitting summaries.}
%%%%%%%%%%%%%%%%%%%%%%%%%%%%%%%%
For $E_1$ versus $\eta$, we fit a straight line in log--log space,
\begin{equation}
    \log E_1 = p\,\log \eta + b,
\end{equation}
using every admissible toy realization. The fit yields
\begin{equation}
    E_1 = C\,\eta^p,
    \qquad
    C = 0.2169272290,
    \qquad
    p = 2.0123679907,
\end{equation}
with 2500 fitted points and $R^2_{\log_{10}}=0.9999196496$.

For $E_0$ versus $\kappa$, we group samples into 14 geometric bins in $\kappa$ and report bin medians with 25\%--75\% quantile bands. A single-$\kappa$ regression leaves substantial scatter ($R^2=0.7668$), whereas two-scale templates that also involve $\eta^2$ track the simulations far more closely ($R^2=0.9738$ for $E_0=A\kappa+B\eta^2$ and $R^2=0.9770$ for $E_0=\sqrt{A\kappa^2+B\eta^4}$), consistent with the interpretation advanced in the main text.

\appsection{Representative Kerr threshold calibration}{app:kerr}

The Kerr band shown in the main text is selected by a small set of quantitative gates on a nearby-spin strip; we record the thresholds here so the construction is reproducible. The broader physical context---near-extremal Kerr quasinormal-mode structure and exceptional-point-adjacent behavior---is summarized in the main text and in Refs.~\cite{YangEtAl2013,YangEtAl2013b,CavalcanteEtAl2024PRL,CavalcanteEtAl2024PRD}.

The label \texttt{pair45} refers to one fixed adjacent-overtone pair in the indexed Kerr QNM outputs used to generate the figures: both members share the same $(l,m)$ and differ by one overtone index in the catalogue ordering \cite{BertiCardosoStarinets2009}. All frequencies $\omega_\pm(a)$ on the strip refer to this pair only.

For a neighboring pair $\omega_\pm(a)$, we use the same local center and half-splitting as elsewhere,
\begin{equation}
    \omega_c(a)=\frac{\omega_+(a)+\omega_-(a)}{2},
    \qquad
    \sigma(a)=\frac{\omega_+(a)-\omega_-(a)}{2},
\end{equation}
together with the spectral compressibility proxy
\begin{equation}
    \eta_K(a;\beta)
    =
    |\sigma(a)|\,T_{\mathrm{spec}}(a;\beta),
    \qquad
    T_{\mathrm{spec}}(a;\beta)=\frac{\beta}{|\Im \omega_c(a)|}.
\end{equation}
On the representative pair45 strip used in the figures, the outer activity hull is the scanned interval
\begin{equation}
    a\in[0.8770,\,0.8810].
\end{equation}
Scanning $\beta\in\{1.5,2.0,3.0\}$ yields the same overlap score on this strip up to an overall rescaling, so we set
\begin{equation}
    \beta_{\mathrm{best}}=2.0.
\end{equation}
The associated outer-hull threshold on the spectral compressibility proxy is then
\begin{equation}
    \eta_* = 0.2457.
\end{equation}
This choice defines the broad $\eta_K$-based candidate band on the activity hull.

Because a same-spin third-pole distance table is not available in the indexed Kerr outputs used for this illustration, we sharpen the band with the local N80 hotspot-cancellation depth as a surrogate isolation statistic,
\begin{equation}
    C_{\mathrm{hotspot},80}(a),
\end{equation}
with working threshold
\begin{equation}
    C_{\mathrm{hotspot},80}(a)\ge 0.01971.
\end{equation}

The combined criterion used in the paper is therefore
\begin{equation}
    \eta_K(a;\beta=2.0) \le 0.2457,
    \qquad
    C_{\mathrm{hotspot},80}(a) \ge 0.01971.
\end{equation}
On this strip the $\eta$-only band covers the full activity hull, whereas the $\eta$-plus-proxy band coincides with the validated support and trims a false-positive width of $0.002000$ in spin.
\bibliographystyle{apsrev4-2}
\bibliography{refall}

\end{document}